\documentclass[pra,twocolumn,amsmath,amssymb]{revtex4}
\usepackage{graphicx}
\usepackage{epsfig}
\newcommand{\bra}[1]{\langle #1 \vert}
\newcommand{\ket}[1]{\vert #1 \rangle}
\newcommand{\braket}[1]{\langle #1 \vert #1 \rangle}
\newcommand{\ketbra}[1]{\vert #1 \rangle\langle #1 \vert}
\begin{document}
\title{Inefficiency and classical communication bounds for conversion between partially entangled pure bipartite states}
\author{Ben Fortescue}
\author{Hoi-Kwong Lo}
\affiliation{%
Center for Quantum Information and Quantum Control,\\
Dept. of Electrical \& Computer Engineering and Dept. of Physics,\\
University of Toronto, Toronto, Ontario, M5S 3G4, CANADA\\
}%
\date{\today}

\begin{abstract}
We derive lower limits on the inefficiency and classical communication costs of dilution between two-term bipartite pure states that
are {\it partially} entangled.
We first calculate explicit relations between the allowable error and classical communication costs of entanglement dilution
using the protocol of \cite{LP} and then consider a two-stage dilution from singlets with this protocol followed
by some unknown protocol for conversion between partially entangled states.  Applying the lower bounds on
classical communication and inefficiency of \cite{HL}, \cite{HW} to this two-stage protocol, we derive
bounds for the unknown protocol.  In addition we derive analogous (but looser) bounds for general pure states.
\end{abstract}
\pacs{03.67.Mn}
\maketitle

\section{Introduction}
In the resource model of quantum information processing, bipartite pure-state entanglement is regarded
as a resource which is fungible (i.e. it can be efficiently converted between different forms) in the
asymptotic limit.  One can concentrate entanglement (convert from many states of low  entanglement to
fewer states of higher entanglement) or dilute entanglement (the reverse process).  The standard unit
of entanglement is the {\it ebit}, the entanglement of an EPR pair, singlet or maximally entangled state (MES) of the form
\begin{equation}
\frac{1}{\sqrt{2}}(\ket{00}+\ket{11})_{AB}
\end{equation}
shared between two parties Alice and Bob.

Asymptotically reversible protocols for concentration and dilution to and from singlets were given by Bennett, Bernstein, Popescu
and Schumacher (BBPS) \cite{BBPS}.  These are reversible in that the entanglement loss due to the process
is only a negligible fraction of the original amount of entanglement - using them one can convert
$M$ copies of an EPR state $\Phi$ to $N$ copies of a pure state $\psi$ such that $\frac{M}{N}\to S(\psi)$ as $N \to\infty$, where
$S(\psi)$ is the entanglement of $\psi$, defined as $-\textrm{Tr}\rho\log_2\rho$ where $\rho=\ketbra{\psi}$.  One can likewise perform the reverse process
with asymptotically negligible loss of entanglement.

However, these protocols may require a non-trivial amount of classical communication. 
While the BBPS protocol for
concentration requires no classical communication, that for dilution requires 2 bits of classical
communication per ebit diluted. If all protocols for dilution required 2 classical bits of
communication per ebit, then entanglement could not truly
be regarded as a fungible resource. For instance, starting
with partially entangled states, superdense coding could
not provide any saving in classical communication cost.
Fortunately, Lo and Popescu \cite{LP} showed that dilution of roughly $NS(\psi)$ singlets
to $N$ copies of a state $\psi$ could be done with only $O(\sqrt{N})$ bits of classical communication
in the large $N$ limit,
compared to $2N$ bits via the BBPS protocol.  It was later shown by Harrow and Lo \cite{HL}
and independently by Hayden and Winter \cite{HW}
that there is a lower bound on the classical communication cost of this dilution of $O(\sqrt{N})$
bits and a lower bound
on the inefficiency of this dilution (i.e. the number of ebits lost in the dilution) of $O(\sqrt{N})$
bits also (so the process can still be
asymptotically efficient with inefficiency per ebit $\to 0$).  Specifically they showed that,
for dilution from $p$ singlets to $N$ copies of a state $\psi$ with probability of success $2^{-s}$
using $c$ bits of classical communication:
\begin{eqnarray}
p&\ge& NS(\psi)+\alpha_{\psi}\sqrt{N}\\
c+s&\ge&\alpha_{\psi}\sqrt{N}
\end{eqnarray}
where for $\ketbra{\psi}$ having eigenvalues $p_i$
\begin{equation}
\alpha_{\psi}^2\equiv\sum_ip_i(\log_2 p_i +S(\psi))^2\qquad(\alpha_{\psi}>0)
\end{equation}
These results all apply solely to dilution from singlets to partially entangled states and do not address
the situation of conversion
between multiple copies of two different partially entangled states, e.g. from roughly $NS(\psi_2)/S(\psi_1)$
copies of an entangled state
$\psi_1$ to $N$ copies of an entangled state $\psi_2$.  Some results regarding dilutions of this kind follow immediately
from the above results -  it is clear from \cite{BBPS} that one can perform such a dilution
asymptotically efficiently, by efficient concentration from the initial state to singlets followed by efficient
dilution from
singlets to the final state.  Likewise there is an upper limit of $O(\sqrt{N})$ bits of classical communication
required for
such a dilution, since one can always concentrate to singlets (at zero classical communication cost) and then dilute
via the Lo-Popescu protocol at a cost of $O(\sqrt{N})$ bits.

However it is not immediately clear what, if any, lower bound applies to the inefficiency and classical communication cost
of converting between partially entangled states - the above results do not forbid there being a cost of e.g. zero
or $O(\log_2 N)$, 
though intuitively
one would expect there to be a non-zero cost in, for example, diluting from a state very close to an EPR state. 

In this paper:
\begin{enumerate}
\item We construct explicit bounds on the classical communication cost
and inefficiency for conversion between partially entangled states.
\item We give explicit examples of partially entangled states
that require $O(\sqrt{N})$ bits of classical communications between
their inter-conversion and likewise for inefficiency.
\item To do so, we have also worked out the dependence of the coefficient of
the $O(\sqrt{N})$ term in the classical communication cost
on the error (as measured in trace distance) in the Lo-Popescu
protocol for entanglement dilution.
\end{enumerate}

\section{Our method}
In this paper, we study the entanglement manipulation process from
roughly $NS(\psi_2)/S(\psi_1)$ copies of a partially entangled
state $\psi_1$ to $N$ copies of another partially entangled state
$\psi_2$. We are interested in its classical communication cost
and also its inefficiency (i.e. loss in entanglement).
Except in the concluding section, we will focus
on the case where $\psi_1$ and $\psi_2$ each have a Schmidt number
of only two. Initially we will focus on the case where $\psi_1$ has
a Schmidt number of only 2.  In other words, $\ket{\psi_1} = a_1\ket{00}+b_1\ket{11}$
and $\psi_2 = a_2\ket{00}+b_2\ket{11}$.

\begin{figure}
\centering
\includegraphics[width=8cm]{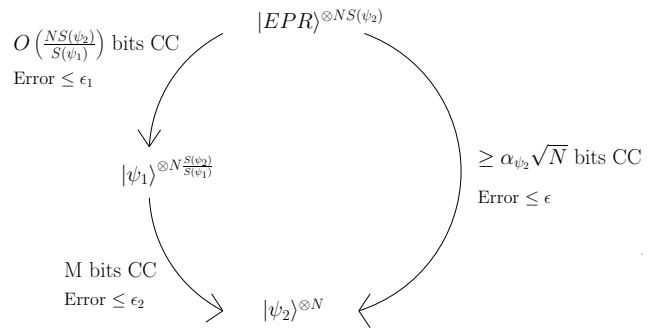}
\caption{Classical communication (CC) costs of entanglement dilution via a two-stage protocol, Lo-Popescu followed by
some unknown protocol for dilution between partially entangled states.  The right-hand side gives the
Harrow-Lo bound on classical communication for the whole process.}\label{fig_dilute}
\end{figure}

The method we use was proposed in \cite{HL} and its idea is to consider a two-stage dilution.  First,
roughly $NS(\psi_2)$ singlets
are diluted to roughly $NS(\psi_2)/S(\psi_1)$ copies of a state $\psi_1$ via the
Lo-Popescu protocol,
using $\beta\sqrt{N}$ bits of classical communication for some $\beta$.  These are then converted to roughly $N$
copies of a state $\psi_2$ via some unknown protocol,
using $m$ bits of classical communication.  This two-stage process is one way of implementing the dilution from
singlets to $\psi_2$ and must therefore obey the lower bound of Harrow-Lo \cite{HL} and Hayden-Winter \cite{HW}.  
The method is illustrated graphically
in Figure \ref{fig_dilute}.
Thus we have that
\begin{eqnarray}
\beta\sqrt{N}+m&\ge& \alpha_{\psi_2}\sqrt{N}\nonumber\\
\therefore m &\ge& (\alpha_{\psi_2}-\beta)\sqrt{N}\label{eq-ccbnd}
\end{eqnarray}
Hence we can derive a lower limit on the classical communication cost of the conversion from $\psi_1$ to $\psi_2$
if we can calculate
the coefficient $\beta$.  This is not derived in \cite{LP} and we do so here.

\section{Classical communication cost in Lo-Popescu}\label{sec-CCLP}
{\it A priori}, the parameter $\beta$ may depend on the target state in question, the allowable error and/or
success probability of the Lo-Popescu protocol.  However the explicit dependence of $\beta$ on these parameters
is not derived in \cite{LP}.  In this section we work out this dependence in detail.

The Lo-Popescu protocol works briefly as follows: $N$ copies of a desired partially entangled state $\psi$ may
be expressed as
\begin{equation}
\psi^N=\Phi^d\otimes\Delta+\epsilon_{LP}\label{eqn-LP}
\end{equation}
where those terms making up $u_1\equiv \Phi^d\otimes\Delta$ have high degeneracy $2^d$ where $d=NS(\psi)-O(\sqrt{N})$
in their Schmidt coefficients
and those in the error term $\epsilon_{LP}$ have low degeneracy.  In diluting from singlets $\Phi$, all that needs to be
teleported is one half of the residual
state $\Delta$ with dimension $2^{O(\sqrt{N})}$ and thus only $O(\sqrt{N})$ bits are required compared to the $2N$
bits that would be needed
to teleport the entire state $\psi^N$ to perform the dilution na\"ively.
\subsection{Error size}
The size of the coefficient on $\sqrt{N}$ in the classical communication cost depends on the allowable error
$\epsilon_{LP}$.  We derive
this dependency for the general two-term case $\ket{\psi}=a\ket{00}+b\ket{11}$, for which we have that the degeneracy of
coefficients in $\ket{\psi^{\otimes N}}$ containing $k$ $a$'s is $N\choose k$.  We define $p=|a|^2$, $q=|b|^2=1-p$ and,
as in \cite{LP}, we define the typical set
as those coefficients for which
\begin{equation}
2^{NS(\psi)-\gamma\sqrt{N}}\le {N\choose k}\le 2^{NS(\psi)+\gamma\sqrt{N}}
\end{equation}
for some coefficient $\gamma$.  We wish to find what range of $k$ this corresponds to.

Using Stirling's approximation for $N!$, we have that
\begin{equation}
\lim_{N\to\infty}{N\choose k}= 2^{NS_b(k/N)}
\end{equation}
where $S_b$ is the binary entropy function
\begin{eqnarray*}
S_b(x)=-x\log_2 x -(1-x)\log_2(1-x)\\
\frac{dS_b}{dx}=-\log_2 x+\log_2(1-x)\\
\frac{dx}{dS_b}=\frac{1}{-\log_2 x+\log_2(1-x)}
\end{eqnarray*}
Hence the typical set corresponds to a range
\begin{equation}
S(\psi)-\gamma/\sqrt{N}\le S_b(k/N)\le S(\psi)+\gamma/\sqrt{N}.
\end{equation}
We consequently calculate $\frac{d(k/N)}{dS_b}=\frac{1}{N}\frac{dk}{dS_b}$
and can thus write $k(S_b)$ as a Taylor expansion to first order in $S_b$ (higher orders are negligible
as $N\to\infty$) about the point $S_b(k/N)=S(\psi)$, $k=Np$.  We find that for the typical
set (assuming henceforth that $p<q$, this is without loss of generality except in the case
$p=q=0.5$ for which the above procedure is not valid but which physically represents having singlets as
the target state in which case no dilution is required)
\begin{equation}
Np - \frac{\gamma\sqrt{N}}{\log_2{q/p}}\le k \le Np + \frac{\gamma\sqrt{N}}{\log_2{q/p}}.
\end{equation}

Thus the typical set is defined by these values of $k$ with the remaining terms
forming the atypical set.  The contribution $\epsilon_{LP_1}$ of the atypical set to the total weight
of the state $\braket{\psi^{\otimes N}}$ is
\[
\sum_{k_{atypical}}{N\choose k}p^kq^{N-k}
\]
a binomial distribution for which in the large $N$ limit we can use the Gaussian approximation
as given in e.g. \cite{Feller}
\begin{equation}
{N\choose k}p^kq^{N-k}\sim \frac{e^{-(k-Np)^2/2Npq}}{\sqrt{2\pi Npq}}
\end{equation}
giving
\begin{align}
\epsilon_{LP_1} &= 2\times{\frac{1}{\sqrt{2\pi Npq}}}\int_{-\infty}^{Np-\frac{\gamma\sqrt{N}}{\log_2(q/p)}}
e^{-\frac{(k-Np)^2}{2Npq}}dk\\
&=2\times\left(1-{\frac{1}{\sqrt{2\pi}}}\int_{-\infty}^{\gamma/\alpha}e^{-x^2/2}dx\right)\label{eqn-atyp}
\end{align}
where $\alpha\equiv\sqrt{pq}\log_2(q/p)=\sqrt{\sum_i p_i[\log_2 p_i +S(\psi)]^2}$, the same $\alpha(\psi)$ defined in \cite{HL}.

We are not aware of any analytical expression for (\ref{eqn-atyp}), but it may be bounded \cite{Feller}
as
\begin{eqnarray}
\sqrt{\frac{2}{\pi}}\left(\frac{1}{(\gamma/\alpha)}-\frac{1}{(\gamma/\alpha)^3}\right)
e^{-\frac{1}{2}\left(\frac{\gamma}{\alpha}\right)^2}
\le \epsilon_{LP_1}\nonumber\\
\le\sqrt{\frac{2}{\pi}}\left(\frac{1}{(\gamma/\alpha)}\right)
e^{-\frac{1}{2}\left(\frac{\gamma}{\alpha}\right)^2}.\label{eqn-errbnd}
\end{eqnarray}

However, the error term $\epsilon_{LP}$ in (\ref{eqn-LP}) does not consist only of atypical terms - as described in
\cite{LP}, in order to ensure a large common degeneracy for the terms in $u_1$ of $2^{NS(\psi)-\omega\sqrt{N}}$ for some
chosen $\omega>\gamma$, certain terms from the typical set need to be assigned to the error term.\\

\noindent {\it Claim One:}
{The contribution  $\epsilon_{{LP}_2}$ to $\braket{\epsilon_{LP}}$, due to
those typical terms which are grouped in the error term in Lo-Popescu's protocol, goes to zero in
the limit that the number of copies N goes to infinity. Hence in this limit $\braket{\epsilon_{LP}}=\epsilon_{LP_1}$}\\

\noindent {\it Proof:} As shown in \cite{LP}, these typical terms have a degeneracy $\le 2^{NS(\psi)-\omega\sqrt{N}}$,
and thus their contribution $\epsilon_{LP_2}$ to $\braket{\epsilon_{LP}}$ satisfies
\begin{equation}
\epsilon_{LP_2}\le 2^{NS(\psi)-\omega\sqrt{N}}\sum_{k_{typical}}p^kq^{N-k}
\end{equation}
Approximating the sum to an integral in the limit of large $N$:
\begin{eqnarray}
\epsilon_{LP_2}&\le& 2^{NS(\psi)-\omega\sqrt{N}}q^N\int_{Np-\frac{\gamma\sqrt{N}}{\log_2(q/p)}}^{Np+\frac{\gamma\sqrt{N}}{\log_2(q/p)}}
(p/q)^kdk\\
&=&\frac{2^{-\omega\sqrt{N}}}{\ln(q/p)}\left[2^{\gamma\sqrt{N}}-2^{-\gamma\sqrt{N}}\right]\\
&\to&\frac{2^{-(\omega-\gamma)\sqrt{N}}}{\ln(q/p)}\textrm{ as }N\to\infty.
\end{eqnarray}

Thus in the asymptotic limit we can take $\braket{\epsilon_{LP}}=\epsilon_{LP_1}$ and
choose $\omega=\gamma+\delta$
for positive $\delta\to 0$, giving a common degeneracy of $2^{NS(\psi)-\gamma\sqrt{N}}$ to the terms
in $u_1$, thus $u_1$ can be expressed as $\Phi^d\otimes\Delta$ where $d=NS(\psi)-\gamma\sqrt{N}$.\\

\noindent {\it Claim Two:} { The error term $\epsilon_{LP_1}$ in Lo-Popescu's protocol is bounded by Eq. {\rm (\ref{eqn-errbnd})}
whose upper bound may be inverted to give
\begin{equation}\label{LPerror}
\frac{\gamma}{\alpha}\le\sqrt{W\left(\frac{2}{\pi\epsilon_{LP_1}^2}\right)}
\end{equation}
where $W()$ is the Lambert W function {\rm \cite{LW}}.}\\

{\it Proof:} It is clear from Eq.(\ref{eqn-errbnd}) that for a given $\epsilon_{LP_1}$, the largest value
of $\gamma/\alpha$ corresponds to the inverse of the upper bound.  The form of the inverse follows
from the definition of $W(x)$ that $W(x)e^{W(x)}=x$.

\subsection{Classical communication cost}
Sch$(u_1)\equiv$ the Schmidt number of $u_1$ is simply $\sum_{k_{typical}}{N\choose k}$.  We are not aware of
any analytical expression for this sum, but it may be bounded \cite{Worsch} as
\begin{equation}
A_02^{NS_b(1/g)}\le\textrm{ Sch}(u_1)\le A_12^{NS_b(1/g)}
\end{equation}
where $1/g=\left(p+\frac{\gamma}{\sqrt{N}\log_2(q/p)}\right)$ and $A_0$ and $A_1$ scale at
most polynomially with $N$.

Thus asymptotically we find
\begin{equation}
\textrm{Sch}(u_1)=A\times 2^{NS(\psi)+\gamma\sqrt{N}}\label{eq-u1dim}
\end{equation}
for some $A$ scaling at most polynomially with $N$.  Since $u_1=\Phi^d\otimes\Delta$ we have that
\begin{eqnarray}
\textrm{Sch}(\Delta)&=&A2^{NS(\psi)+\gamma\sqrt{N}-d}\nonumber\\
&=&A\times 2^{2\gamma\sqrt{N}}\label{eq-deltadim}
\end{eqnarray}

It was shown in \cite{Lo} that production of an entangled state via quantum teleportation can be performed with
half the classical communication cost of na\"ive teleportation of the state $\Delta$.  
Hence the asymptotic classical communication cost of diluting from singlets to $N$ copies of a state $\psi$ via
Lo-Popescu is $\frac{1}{2}\times 2\times2\gamma\sqrt{N}=2\gamma\sqrt{N}$ bits.

\section{Relation between protocol errors}\label{sec-errs}
We first define the trace distance
between states with density matrices $\sigma$ and $\rho$, $D(\sigma,\rho)\equiv\textrm{Tr}|\sigma-\rho|$.\\
In our two-stage protocol, we start with roughly $NS(\psi_2)$ singlets with an overall density matrix $\sigma$. 
We then dilute
using Lo-Popescu to a state $\Lambda'$ which is close to the desired state of $NS(\psi_2)/S(\psi_1)$
copies of $\psi_1$, with density matrix $\Lambda$.  Finally we dilute $\Lambda'$ via some unknown
protocol to $\rho''$, close to the desired state $\rho$ of $N$ copies of a state $\psi_2$.  $\rho''$
differs from $\rho$ due to the errors introduced at both dilution stages - we in addition define $\rho'$ as the
state which one would acquire from the unknown protocol if $\Lambda$ was used as the input state.

\begin{figure}
\includegraphics[width=8cm]{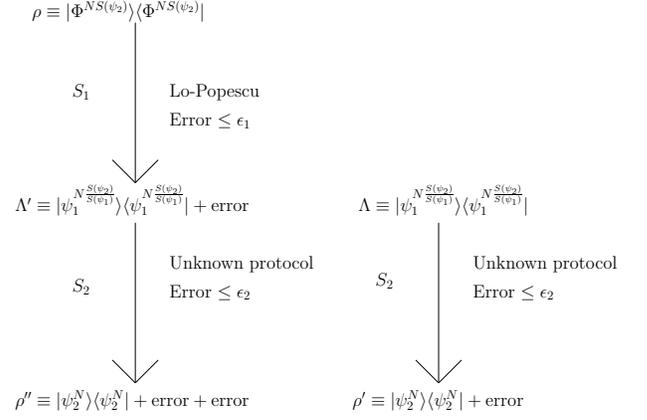}
\caption{Errors for different stages of the two-stage dilution}\label{fig-errors}
\end{figure}

We define the upper bounds of the errors in the protocols as follows:
\begin{eqnarray*}
D(\Lambda,\Lambda')&\le& \epsilon_1\textrm{ (error in Lo-Popescu)}\\
D(\rho',\rho)&\le& \epsilon_2 \textrm{ (error in unknown protocol)}\\
D(\rho'',\rho)&\le&\epsilon\textrm{ (overall error in two-stage protocol)}
\end{eqnarray*}
The above is illustrated in Figure \ref{fig-errors}.

Since our final bounds will depend only on $\epsilon_2$ we are free to choose our allowable
error $\epsilon_1$ in the Lo-Popescu dilution as long as it is consistent with the requirement
of the Harrow-Lo bound (under ``Definition of error'' in Theorem 1 of \cite{HL}) that
\begin{equation}
D(\rho,\rho'')\le 0.01\label{eq-hlbnd}
\end{equation}

We can describe the Lo-Popescu dilution as a trace-preserving operation $S_1$, likewise
the unknown protocol as a trace-preserving operation $S_2$, so that
\begin{eqnarray*}
\Lambda'&=&S_1\sigma\\
\rho''&=&S_2\Lambda'=S_2S_1\sigma\\
\rho'&=&S_2\Lambda
\end{eqnarray*}
From the triangle inequality (see for example \cite{NC}), we have that:
\begin{eqnarray*}
\epsilon_2=D(\rho'',\rho)&=&D(S_2S_1\sigma,\rho)\\
&\le& D(S_2S_1\sigma, S_2\Lambda)+D(S_2\Lambda,\rho)\\
&\le& D(S_1\sigma,\Lambda)+D(\Lambda,\rho)\\
&=&D(S_1\sigma,\Lambda)+\epsilon_2
\end{eqnarray*}
since trace-preserving quantum operations are contractive (applying $S_2$ cannot increase the
trace distance).
\begin{eqnarray*}
D(S_1\sigma,\Lambda)+\epsilon_2 &=&  D(\Lambda,\Lambda')+\epsilon_2\\
&=& \epsilon_1+\epsilon_2
\end{eqnarray*}
hence
\begin{equation}
D(\rho,\rho'')\le\epsilon_1+\epsilon_2.\label{eq-errs}
\end{equation}

In performing a two-stage dilution,we are free to choose the $\epsilon_1$ for our Lo-Popescu protocol.
However in order to derive inefficiency and classical communication bounds for the second stage
we must be able to apply the Harrow-Lo bound and hence must satisfy (\ref{eq-hlbnd}).  But given this
restriction, we would like $\epsilon_1$ to be as large as possible so that $\gamma/\alpha_{\psi_1}$
is small and hence the inefficiency and classical communication bounds are restrictive; $\epsilon_1\to 0$ would
give $\gamma/\alpha\to\infty$ and hence no meaningful bound (since the Lo-Popescu classical
communication cost and inefficiency would then be very large for almost all $\psi_1$ and hence satisfy the Harrow-Lo bound
without the need for any classical communication or inefficiency from the second stage of the dilution). 
Thus for given $\epsilon_2<0.01$, from (\ref{eq-errs}) we satisfy (\ref{eq-hlbnd}) by
setting
\begin{equation}
\epsilon_1=0.01-\epsilon_2
\end{equation}

In Lo-Popescu the (normalized) state obtained is $\ketbra{\Phi^d\otimes\Delta+\epsilon_{LP}}$
and the (normalized) desired state is $\frac{\ketbra{\Phi^d\otimes\Delta}}{1-\epsilon_{LP_1}}$.
Thus we find, from Claim 1:
\begin{eqnarray}
\textrm{Tr}|\Lambda'-\Lambda|&=
\textrm{Tr}&\Bigg|\frac{\ketbra{\Phi^d\otimes\Delta}}{1-\epsilon_{LP_1}}\nonumber\\
&&-\ketbra{\Phi^d\otimes\Delta+\epsilon_{LP}}\Bigg|\\
&=\textrm{Tr}&\Bigg|\frac{\epsilon_{LP_1}\ketbra{\Phi^d\otimes\Delta}}{1-\epsilon_{LP_1}}\nonumber\\
&&-\ket{\Phi^d\otimes\Delta}\bra{\epsilon_{LP}}\nonumber-\ket{\epsilon_{LP_1}}\bra{\Phi^d\otimes\Delta}\nonumber\\
&&-\ketbra{\epsilon_{LP}}\Bigg|\\
&=\textrm{Tr}&\left|\frac{\epsilon_{LP_1}\ketbra{\Phi^d\otimes\Delta}}{1-\epsilon_{LP_1}}
-\ketbra{\epsilon_{LP}}\right|\nonumber\\
&=&|\epsilon_{LP_1}|+|-\epsilon_{LP_1}|=2\epsilon_{LP_1}\le\epsilon_1
\end{eqnarray}

since $\braket{\epsilon_{LP}}=\epsilon_{LP_1}$ asymptotically.  Thus for a chosen allowable error $\epsilon_1$, $\epsilon_{LP_1}\le\frac{\epsilon_1}{2}$.
Hence from Claim 2:
\begin{equation}
\gamma/\alpha\le\sqrt{W\left(\frac{8}{\pi(0.01-\epsilon_2)^2}\right)}
\end{equation}
and so for a given stage 2 error $\epsilon_2\le0.01$ we can begin our two-stage dilution with a Lo-Popescu dilution
with a classical communication cost no greater than
\begin{equation}
2\sqrt{W\left(\frac{8}{\pi(0.01-\epsilon_2)^2}\right)}\alpha_{\psi_1}\sqrt{\frac{NS(\psi_2)}{S(\psi_1)}}\textrm{ bits.}\label{eq-lpcc}
\end{equation}

\begin{figure}
\centering
\includegraphics[angle=270,width=8cm]{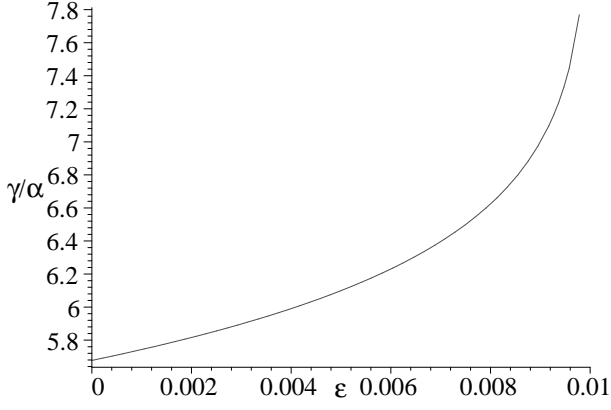}
\caption{Upper limit on $\gamma/\alpha$ as a function of $\epsilon_2$}\label{fig-gaplot}
\end{figure}

A plot of the upper bound on $\gamma/\alpha$ noted in Claim 2 against $\epsilon_2$ is given in Figure \ref{fig-gaplot}.

Applying the bound in \cite{HL} and using Eq. (\ref{eq-ccbnd}) and Eq. (\ref{eq-lpcc})
\begin{equation}
2\sqrt{W\left(\frac{8}{\pi(0.01-\epsilon_2)^2}\right)}\alpha_{\psi_1}\sqrt{\frac{NS(\psi_2)}{S(\psi_1)}}+m\ge\alpha_{\psi_2}\sqrt{N}
\end{equation}
\begin{eqnarray}
m \ge\left(\alpha_{\psi_2}-2\sqrt{W\left(\frac{8}{\pi(0.01-\epsilon_2)^2}\right)}\alpha_{\psi_1}\sqrt{\frac{S(\psi_2)}{S(\psi_1)}}\right)&\sqrt{N}\nonumber\\
\textrm{ bits.}&
\end{eqnarray}

As one would expect, this bound is strictest for a dilution protocol with $\epsilon_2=0$, for which
we have
\begin{equation}
m \ge\left(\alpha_{\psi_2}-5.68\alpha_{\psi_1}\sqrt{\frac{S(\psi_2)}{S(\psi_1)}}\right)\sqrt{N}\textrm{ bits.}
\end{equation}
where the figure $5.68$ is to two decimal points.

\section{The inefficiency bound}\label{sec-ineff}
We can do an analogous calculation of the inefficiency in dilution between partially
entangled states by considering the same two-stage dilution as before and the ebits lost
at each stage, since we know that (ebits lost in Lo-Popescu dilution from
singlets to $\psi_1$)+(ebits lost in dilution from $\psi_1$ to $\psi_2$) $\ge$ the lower bound
on inefficiency given in \cite{HL} $=\alpha_{\psi_2}\sqrt{N}$ ebits.

\subsection{Inefficiency in Lo-Popescu}
In the Lo-Popescu protocol the only state that needs to be teleported is $\Delta$ and the inefficiency
arises in this teleportation, which requires $\log_2$(Sch($\Delta$)) ebits to teleport but which
only provides $S(\Delta)$ ebits of shared entanglement.

We find that the entropy of those terms in $\epsilon_{LP_2}$ is asymptotically
negligible (proportional to $2^{-(\omega-\gamma)\sqrt{N}})$, and hence that the entropy of $u_1$ is asymptotically just that of the typical set.
\begin{eqnarray}
S(u_1)&=&-\sum_{k_{typical}}{N\choose k}p^kq^{N-k}\log_2(p^kq^{N-k})\nonumber\\
&\approx&\frac{-1}{\sqrt{2\pi Npq}}
(N\log_2 q)\int_{Np-\frac{\gamma\sqrt{N}}{\log_2(q/p)}}^{Np+\frac{\gamma\sqrt{N}}{\log_2(q/p)}}e^{-(k-Np)^2/2Npq}dk\nonumber\\
&=&\frac{-1}{\sqrt{2\pi}}\Bigg[N\log_2 q \int_{-\gamma/\alpha}^{\gamma/\alpha}e^{-x^2/2}dx\nonumber\\
&&+Np\log_2(p/q)\int_{-\gamma/\alpha}^{\gamma/\alpha}e^{-x^2/2}dx\nonumber\\
&&+ \sqrt{N}pq\int_{-\gamma/\alpha}^{\gamma/\alpha}xe^{-x^2/2}dx\Bigg]\nonumber\\
&=&\frac{NS(\psi)}{\sqrt{2\pi}}\int_{-\gamma/\alpha}^{\gamma/\alpha}e^{-x^2/2}dx\nonumber\\
&=&NS(\psi)[1-\epsilon_{LP_1}],
\end{eqnarray}
so the entanglement of normalized $u_1$ is asymptotically just $NS(\psi)$.
$u_1=\Phi^d\otimes\Delta$, thus the entanglement of $\Delta$ is
$NS(\Psi)-d=\gamma\sqrt{N}$.  So from equation (\ref{eq-deltadim})
the asymptotic inefficiency (the number of ebits lost) in diluting
from singlets to $N$ copies of $\psi$ via Lo-Popescu is
\begin{equation}
\log_2(\textrm{Sch}(\Delta))-S(\Delta)=2\gamma\sqrt{N}-\gamma\sqrt{N}=\gamma\sqrt{N}\textrm{ ebits}
\end{equation}

Hence for the two-stage dilution, as described in section \ref{sec-errs}, we can always
choose our error in Lo-Popescu such that $\gamma/\alpha\le\sqrt{W\left(\frac{8}{\pi(0.01-\epsilon_2)^2}\right)}$
and so from the Harrow-Lo inefficiency bound we have that, for a loss of $k$ ebits from the unknown protocol
\begin{eqnarray}
\sqrt{W\left(\frac{8}{\pi(0.01-\epsilon_2)^2}\right)}\alpha_{\psi_1}+k\ge\alpha_{\psi_2}\sqrt{N}\quad&\\
k \ge\left(\alpha_{\psi_2}-\sqrt{W\left(\frac{8}{\pi(0.01-\epsilon_2)^2}\right)}\alpha_{\psi_1}\right)\sqrt{N}\textrm{ ebits.}\quad&
\end{eqnarray}
with the strictest bound again for an error $\epsilon_2=0$, with a minimum inefficiency of $2.64\alpha_{\psi_2}\sqrt{N}$ (to two decimal points) ebits, half
of that in the classical communication cost bound.

\section{The general pure state case}
For the general pure state
\begin{equation}
\ket{\psi}=\sum_{i=1}^m\sqrt{p_i}\ket{ii}
\end{equation}
we can perform an analogous analysis to the single-qubit case, bounding the degeneracy of terms in $\psi^N$
\begin{equation}
NS(\psi)-\gamma\sqrt{N}\le \log_2{N\choose k_1\ldots k_m}\le NS(\psi)+\gamma\sqrt{N}
\end{equation}

We use the substitution (from \cite{Arenbaev}, though it mistakenly quotes $\pi_i^{-1}$ instead of $\pi_{i-1}^{-1}$ in the numerator) of
\begin{equation}
y_i=\frac{k_i-(N-k_i-\ldots-k_{i-1})p_i\pi_{i-1}^{-1}}{\sqrt{Np_i\pi_i\pi_{i-1}^{-1}}},\quad\pi_i=1-p_1-\ldots p_i
\end{equation}
giving the asymptotic approximation
\begin{equation}
{N\choose k_1\ldots k_m}p_1^{k_1}\ldots p_m^{k_m} \sim \frac{e^{-(y_1^2+\ldots+y_{m-1}^2)/2}}{(\sqrt{2\pi N})^{m-1}
\sqrt{p_1p_2\ldots p_m}}\label{Pmulti}
\end{equation}

Differentiating the degeneracies with respect to $y_i$, we find that for the typical set
\begin{equation}
\left|\sum_{i=1}^{m-1} \Omega_i\Delta y_i \right|\le \gamma 
\end{equation}
where
\begin{equation}
\Omega_i=\sqrt{\frac{p_i}{\pi_i\pi_{i-1}}}\sum_{j=i+1}^m p_j\log_2(p_j/p_i)\label{Omegarange}.
\end{equation}

We are unaware of any expression for the integral of
the probability distribution (\ref{Pmulti}) over the range (\ref{Omegarange}).  Instead we choose
an expression $\Omega_t\ge |\Omega_i|\textrm{ }\forall\textrm{ } i<m$ and integrate (\ref{Pmulti}) outside the range $\pm\gamma/\Omega_{max}$ for all $y_i$
to obtain an upper bound on $\epsilon_{LP_1}$.  Doing so we find
\begin{eqnarray}
\epsilon_{LP_1}&\le&2\left(\frac{1}{\sqrt{2\pi}}\int_{-\frac{\gamma}{\Omega_t}}^{\frac{\gamma}{\Omega_t}}e^{-y^2/2}dy\right)^{m-1}\\
&\le& \frac{2}{[\sqrt{2\pi}(\gamma/\Omega_t)]^{m-1}}e^{-(m-1)\left(\frac{\gamma}{\Omega_t}\right)^2/2}
\end{eqnarray}
which inverts to
\begin{equation}
\frac{\gamma}{\Omega_t}\le\sqrt{W\left(\frac{1}{2\pi(\epsilon_{LP_1}/2)^{\frac{2}{m-1}}}\right)}\label{error-multi}
\end{equation}

\subsection{Inefficiency and classical communication bounds for the general pure state}
The remainder of the analysis follows as for the single-qubit case - in particular we find that Claims
1 and 2 still apply (with (\ref{error-multi}) substituted into Claim 2) and our final results are, that for an
$m$-bit classical communication cost and $k$-ebit inefficiency diluting from $\psi_1$ to $\psi_2$, where $\psi_1$
has Schmidt number $m$\\
\begin{align}
m&\ge\left(\alpha_{\psi_2}-2\sqrt{W\left(\frac{1}{2\pi(\frac{0.01-\epsilon_2}{4})^{\frac{2}{m-1}}}\right)}
{\Omega_t}_{\psi_1}\sqrt{\frac{S(\psi_2)}{S(\psi_1)}}\right)\sqrt{N}\\
k&\ge\left(\alpha_{\psi_2}-\sqrt{W\left(\frac{1}{2\pi(\frac{0.01-\epsilon_2}{4})^{\frac{2}{m-1}}}\right)}
{\Omega_t}_{\psi_1}\sqrt{\frac{S(\psi_2)}{S(\psi_1)}}\right)\sqrt{N}
\end{align}
\section{Conclusion}
We have derived lower bounds for the classical communication cost and inefficiency of converting between
pure bipartite entangled states expressible in two terms.  Specifically, we have proven the following results:\\

\noindent {\it Claim 3: }For any protocol which converts via LOCC $NS(\psi_2)/S(\psi_1)$ copies
of a state $\psi_1$ into $N$ copies of state $\psi_2$ with at most error $\epsilon_2< 0.01$,
in the limit of large $N$ and where $\psi_1, \psi_2$ are bipartite pure entangled states and $\psi_1$ is 
expressible in two terms,
the classical communication cost is at least
\begin{equation}
\left(\alpha_{\psi_2}-2\sqrt{W\left(\frac{8}{\pi(0.01-\epsilon_2)^2}\right)}\alpha_{\psi_1}\sqrt{\frac{S(\psi_2)}{S(\psi_1)}}\right)\sqrt{N}\textrm{ bits}.
\end{equation}
\\

\noindent {\it Claim 4: }{For the protocol described in Claim 3, the inefficiency
is at least
\begin{equation}
\left(\alpha_{\psi_2}-\sqrt{W\left(\frac{8}{\pi(0.01-\epsilon_2)^2}\right)}\alpha_{\psi_1}\sqrt{\frac{S(\psi_2)}{S(\psi_1)}}\right)\sqrt{N}\textrm{ ebits}.
\end{equation}
}
{\it Proof: } As shown in Sections \ref{sec-CCLP}, \ref{sec-errs} and \ref{sec-ineff}.

These bounds are meaningful in that they give non-zero
classical communication costs for conversions between possible states, however they only give a partial
ordering on entangled states due to the asymmetry of the coefficients involved.  Namely, nonzero classical communication
is required for entanglement conversion from $\psi_1$ to $\psi_2$ when
\begin{equation}
\frac{\alpha_{\psi_2}}{S(\psi_2)}\ge
2\sqrt{W\left(\frac{8}{\pi(0.01-\epsilon_2)^2}\right)}\frac{\alpha_{\psi_1}}{S(\psi_1)}
\end{equation}
and nonzero inefficiency when
\begin{equation}
\frac{\alpha_{\psi_2}}{S(\psi_2)}\ge
\sqrt{W\left(\frac{8}{\pi(0.01-\epsilon_2)^2}\right)}\frac{\alpha_{\psi_1}}{S(\psi_1)}
\end{equation}
So, for example, for a protocol with essentially zero error, there will be a non-zero classical
communication cost in diluting e.g. states $\ket{\psi_1}=\sqrt{0.43}\ket{00}+\sqrt{0.57}\ket{11}$
to states $\ket{\psi_2}=\sqrt{0.14}\ket{00}+\sqrt{0.86}\ket{11}$, and a non-zero inefficiency
in diluting the same $\psi_1$ to states $\ket{\psi_2}=\sqrt{0.3}\ket{00}+\sqrt{0.7}\ket{11}$.  I.e. these bounds do give useable limits on allowable dilutions.

We have also derived explicit expressions for various details of the Lo-Popescu protocol, in
particular the dependence of the coefficient in the $O(\sqrt{N})$ classical communication cost on the allowable error.

There is however no reason to believe that these bounds cannot be improved - their derivation is dependent on considering
one particular entanglement dilution protocol (Lo-Popescu) and a tighter bound could potentially be derived
by considering a different protocol with lower inefficiency and classical communication costs.

\subsection{General pure states}

For dilution of general pure states, we have found:\\

\noindent {\it Claim 5. }For any protocol which converts via LOCC $NS(\psi_2)/S(\psi_1)$ copies
of a state $\psi_1$ into $N$ copies of state $\psi_2$ with at most error $\epsilon_2< 0.01$,
in the limit of large $N$ and where $\psi_1, \psi_2$ are bipartite pure entangled states and $\psi_1$ is of Schmidt number $b$,
the classical communication cost is at least
\begin{equation}
\left(\alpha_{\psi_2}-2\sqrt{W\left(\frac{1}{2\pi(\frac{0.01-\epsilon_2}{4})^{\frac{2}{b-1}}}\right)}{\Omega_t}_{\psi_1}\sqrt{\frac{S(\psi_2)}{S(\psi_1)}}\right)\sqrt{N}\textrm{ bits}.
\end{equation}
for an ${\Omega_t}_{\psi_1} \ge |{\Omega_i}_{\psi_1}|\textrm{ }\forall\textrm{ }i<b$.
\\

\noindent {\it Claim 6. } For the protocol described in Claim 5, the inefficiency
is at least
\begin{equation}
\left(\alpha_{\psi_2}-\sqrt{W\left(\frac{1}{2\pi(\frac{0.01-\epsilon_2}{4})^{\frac{2}{b-1}}}\right)}{\Omega_t}_{\psi_1}\sqrt{\frac{S(\psi_2)}{S(\psi_1)}}\right)\sqrt{N}\textrm{ ebits}.
\end{equation}

\subsubsection{Choosing $\Omega_t$}
Still undefined is how one chooses $\Omega_t$.  Ideally one would would take $\Omega_t$ equal to the
largest $\Omega_i$, but it can be seen from (\ref{Omegarange}) that the values of $\Omega_i$ are
dependent on the arbitrary ordering of the coefficients $p_i$.  Thus the tightest bound
this analysis provides is for $\Omega_t$ equal to the maximum $\Omega_i$ for the ordering
in which this maximum is smallest.  For a given state this is certainly a well-defined quantity,
but we do not have a general formula for it in terms of the $p_i$'s.

An alternative prescription for $\Omega_t$ is to order $p_1\le p_2\le\ldots \le p_b$ and use
\begin{equation}
\Omega_t=\sqrt{\frac{p_{b-1}}{\pi_{b-1}\pi_{b-2}}}\sum_{i=2}^b p_i\log(p_i/p_1)\label{Omegadefined}
\end{equation}
which is simply expressed, but depending on the state may provide a much looser bound.  Whatever
the chosen $\Omega_t$, using a single bound for all the $y_i$ clearly makes the overall bound looser
than in the two-Schmidt term case.

However even the general $\Omega_t$ in (\ref{Omegadefined}) can give specific bounds on
classical communication and inefficiency.  E.g. we find that for an error-free
protocol diluting from many copies of $\sqrt{0.3}\ket{11}+\sqrt{0.3}\ket{22}+\sqrt{0.4}\ket{33}$
to $\sqrt{0.1}\ket{11}+\sqrt{0.1}\ket{22}+\sqrt{0.8}\ket{33}$ at least 0.29$\sqrt{N}$ bits
of classical communication are required and at least 0.87$\sqrt{N}$ ebits are lost.  It may well
be feasible to obtain much tighter bounds in specific cases.


\begin{thebibliography}{99}
\bibitem{LP}H. K. Lo, S. Popescu,
arXiv:quant-ph/9912009, Phys. Rev. Lett. \textbf{83}, 1459 (1999).
\bibitem{HL}A. Harrow, H. K. Lo,
arXiv:quant-ph/0204096, IEEE Trans. Inf. Th. Vol. 50, \textbf{2}, 319 (2004),
\bibitem{HW}
P. Hayden, A. Winter,
arXiv:quant-ph/0204092, Phys. Rev. A \textbf{67}, 012326 (2003)
\bibitem{BBPS}C. H. Bennett, H. J. Bernstein, S. Popescu and B. Schumacher,
arXiv:quant-ph/9511030, Phys. Rev. A \textbf{53}, 2046 (1996).
\bibitem{Feller}W. Feller
{\it An Introduction to Probability Theory and Its Applications}
(John Wiley \& Sons, 1957) p.170
\bibitem{LW}R. Corless, G. Gonnet, D. Hore, D. Jeffrey and D. Knuth,
Adv. Comput. Math. \textbf{5}, 329 (1996)
\bibitem{Worsch}T. Worsch,
Technical Report 31/94, Universit{\"a}t Karlsruhe, Fakult{\"a}t f{\"u}r Informatik (1994)
\bibitem{Lo}H. K. Lo,
arXiv:quant-ph/9912009, Phys. Rev. A \textbf{62}, 012313 (2000)
\bibitem{NC}M. Nielsen and I. Chuang,
{\it Quantum Computation and Quantum Information}
(Cambridge University Press, 2000) p.406
\bibitem{Arenbaev}N.K. Arenbaev
{\it Asymptotic behaviour of the multinomial distribution}
Theory of Probability and its Applications {\bf 4} Vol XXI (1976).
\end{thebibliography}
\end{document}